# Hardness of FeB$_4$: Density functional theory investigation

Miao Zhang[1,*], Mingchun Lu[2], Yonghui Du[1], Lili Gao[1], Cheng Lu[3], and Hanyu Liu[4,*]

[1]*Department of Physics, Beihua University, Jilin 132013, China*

[2]*Department of Aeronautical Engineering Professional Technology, Jilin Institute of Chemical Technology, Jilin 132102, China*

[3]*Department of Physics, Nanyang Normal University, Nanyang 473061, China*

[4]*Department of Physics and Engineering Physics, University of Saskatchewan, Saskatchewan, Canada, S7N 5E2*

A recent experimental study reported the successful synthesis of an orthorhombic FeB$_4$ with a high hardness of 62(5) GPa [Gou *et al*., Phys. Rev. Lett. 111,157002 (2013)], which has reignited extensive interests on whether transition-metal borides (TRBs) compounds will become superhard materials. However, it is contradicted with some theoretical studies suggesting transition-metal boron compounds are unlikely to become superhard materials. Here, we examined structural and electronic properties of FeB$_4$ using density functional theory. The electronic calculations show the good metallicity and covalent Fe-B bonding. Meanwhile, we extensively investigated stress-strain relations of FeB$_4$ under various tensile and shear loading directions. The calculated weakest tensile and shear stresses are 40 GPa and 25 GPa, respectively. Further simulations (e.g. electron localized function and bond length along the weakest loading direction) on FeB$_4$ show the weak Fe-B bonding is responsible for this low hardness. Moreover, these results are consistent with the value of Vickers hardness (11.7－32.3 GPa) by employing different empirical hardness models and below the superhardness threshold of 40 GPa. Our current results suggest FeB$_4$ is a hard material and unlikely to become superhard (> 40 GPa).

## I. INTRODUCTION

Superhard materials have attracted extensive attentions due to their widely technological applications in cutting and polishing tools, coatings and abrasives, etc. Traditionally, it is commonly accepted that superhard materials (diamond, cubic BN, $BC_2N$, etc.) are formed by light elements (B, C, N and O), because they contain short and strong covalent bonds providing to resist both elastic and plastic deformations. Besides light covalent element compounds, another family of materials consisting of heavy transition-metals (W, Re, Os, Ru, etc.) and light elements are also considered to be potential superhard materials, where heavy transition-metal and light elements provide high valence electron densities and strong covalent bonds, respectively.[1-4] Recently, many transition-metal light-element compounds have been successfully synthesized, such as $OsB_2$,[1] $ReB_2$,[5-10] $RuB_2$,[8] $WB_4$,[8,11-13] and $CrB_4$,[4] etc. These transition-metal boride compounds have attracted special attention because of high hardness due to their high boron content.[4,14,15] However, many theoretical researches[8,9, 16-22] indicate that these transition-metal boride compounds should not become superhard materials because of their low calculated values of Vickers hardness.

Very recently, Gou *et al.*[14] reported that a highly incompressible $FeB_4$ was synthesized with ultra-high nanoindentation hardness ranging from 43 to 70 GPa (an average value of ~62 GPa), opening a new avenue to synthesize superhard transition-metal light element compounds with advanced mechanical properties. This structure consists of boron cages surrounding each Fe atom connected by strong B-B bonds forming a three-dimensional boron network, where Fe provides high valence electron density. This helps to resist both elastic and plastic deformations, which may be the origin of ultra-high hardness in $FeB_4$. This $FeB_4$ compound was found to have a high (Vickers) hardness exceeding 40 GPa, comparable to c-BN. Interestingly, according to previous theoretical predictions, the bulk and shear modulus $FeB_4$ are only almost half those of c-BN.[4] This suggests the hardness of $FeB_4$ should be around 30 GPa. Generally, the hardness is strongly correlated to bulk and shear modulus, since the hardness is deduced from the size of the indentation after deformation. A hard material typically requires a high bulk modulus to support the volume decrease created by the applied pressure, and a high hardness requires limiting the creation and mobility of dislocations, which is largely governed by the resolved shear stress of the material. The bulk and shear modulus has been considered as very important parameters, governing the indentation hardness. The estimated hardness value (~30 GPa) is lower than superhard criteria (>40 GPa) and arises the question correspondingly: whether $FeB_4$ belongs to a superhard material and if it does, what is the origin of its superhard nature? Therefore, it is necessary to explore the mechanical properties of $FeB_4$, which is crucial to understanding its hardness. For this purpose, we studied stress-strain relations of $FeB_4$ under various tensile and shear loading directions, which are helpful to improve our understanding of its superhard nature. These results show the weak Fe-B bonding is responsible for this low hardness. Moreover, the ideal strength results are consistent with the value of Vickers hardness (11.7－32.3 GPa) by employing different empirical hardness models. These results suggest $FeB_4$ is a hard material with a low hardness value of ~25 GPa. The present results shed strong light on the critical role played by ideal strength in proper understanding of hard nature of transition-metal boride compounds.

## II. COMPUTATIONAL METHOD

Recent advances in computational physics have made it possible to calculate the stress-strain relations of a perfect crystal in various tensile and shear deformation directions under the normal compressive pressure beneath an indenter.[23-28] In this work, we report a first study of the ideal strength of $FeB_4$ using first-principles electronic structure calculations. The underlying *ab initio* structural relaxations and the calculations of ideal tensile and shear strength were carried out using the density functional theory with the Perdew-Burke-Ernzerh generalized gradient approximation[29] exchange-correlation potential as implemented in the VASP code.[30] The all-electron projector-augmented wave (PAW) method[31] was adopted with $3p^63d^74s^1$ and $2s^22p^1$ treated as valence electrons for Fe and B, respectively. The cutoff energy of 450 eV for the expansion of the wave function into plane waves and Monkhorst-Pack[32] $k$-points were chosen to 5×5×8 in the Brillouin zone for FeB$_4$ to ensure that all the calculations are well converged. Elastic constants were calculated by the strain-stress method, and bulk modulus and shear modulus were thus derived from the Voigt-Reuss-Hill averaging scheme.[33,34]

## III. RESULTS AND DISCUSSION

In Table I, we listed the optimized structural lattice constants of FeB$_4$ at ambient pressure, and made a comparison with previously reported theoretical and experimental data, which yields a good agreement.[35] In fact, calculated elastic constants not only are helpful to understand the mechanical properties but also provide very useful information to estimate the hardness of materials. Toward this goal, we studied the mechanical properties of FeB$_4$ by means of the strain-stress method in combination with first-principles density functional calculations. The calculated bulk modulus of 265 GPa is in agreement with previous theoretical studies[36], suggesting the ultra-incompressible nature of FeB$_4$. According to the mechanical stability criteria[37] of orthorhombic phase, calculated elastic constants show FeB$_4$ is mechanically stable under ambient condition. It is noteworthy that $B_0$ of FeB$_4$ is ~57% and ~66% of those found in diamond and c-BN, respectively. Moreover, the shear modulus ($G$) of FeB$_4$ was calculated to be 197.97 GPa (Table I), which is only 36% and 48% of those in diamond and c-BN, respectively. In fact, the hardness of a material is strongly correlated to the bulk modulus and shear modulus, since the hardness of a crystal is the ability to resist plastic deformation from hydrostatic compression, tensile load and shear. Therefore, the hardness of FeB$_4$ should be around 48%-66% of those in c-BN. In order to investigate the Vickers hardness ($H_v$) of FeB$_4$, we estimate the hardness of FeB$_4$ at ambient pressure employing an empirical model.[40-45] The predicted hardness for FeB$_4$ is ~11-32 GPa (Table I), which is half that of the experimental value and much less than those found in diamond (~90 GPa) and c-BN (~65 GPa). Interestingly, these results indicate FeB$_4$ is not as superhard as suggested by experiment.[23]

We have extensively examined stress-strain relations of FeB$_4$ under tensile loading,

as shown in Fig. 1. It is seen that $FeB_4$ has strong stress responses in the <001>, <011>, and <010> directions with the peak tensile stresses between 50 and 65 GPa. The weakest tensile direction is along <111> direction, indicating $FeB_4$ would likely experience cleavage in the <111> direction with an ideal tensile strength of 40.4 GPa. In the meantime, it also exhibits comparable weaker anisotropy in its peak tensile stresses, with descending magnitudes for <100>, <001>, <011>, <110>, <010>, and <111> directions at the ratio of 1.13:1.49:1.32:1.18:1.41:1.00, compared to 2.40:1.35:1.00 for diamond and 3.00:1.45:1.00 for c-BN in <100>, <110>, and <111> directions.[24] At the critical tensile strain ($\varepsilon = 0.23$), the stress has only decreased slightly from the peak value. The corresponding structural snapshot indicates a strong $sp^3$ bonding character of sudden "hard" breaking at $\varepsilon = 0.24$. We now turn to a detailed analysis of atomistic structural deformation modes in $FeB_4$ to examine microscopic mechanism of its ideal strength and the fracture behavior under the tensile loading conditions. The bond length as a function of the <111> tensile direction strain is shown in Fig. 2. The bond-lengths of $Fe_1$-$B_1$, $Fe_3$-$B_5$, $Fe_2$-$B_2$ and $Fe_2$-$B_6$ are 2.247 Å, 2.247 Å, 2.124 Å and 2.124 Å at equilibrium ($\varepsilon = 0$), respectively. With increasing the <111> tensile deformation, bond-lengths of $Fe_1$-$B_1$ ($Fe_3$-$B_5$) and $Fe_2$-$B_2$ ($Fe_2$-$B_6$) gradually increase. When tensile strain achieves 0.23, their bond-lengths become 2.862 Å ($Fe_1$-$B_1$ and $Fe_3$-$B_5$) and 2.811 Å ($Fe_2$-$B_2$ and $Fe_2$-$B_6$), which are shown in the right panel of Fig. 2. On the other hand, bond-lengths of $B_3$-$B_4$ and $B_7$-$B_8$ are also increased from 1.839 Å ($\varepsilon = 0$) to 2.142 Å. Note that B-B bonds remain strong up to the <111> tensile. Once the tensile strain achieves 0.24, bonds of $B_3$-$B_4$ and $B_7$-$B_8$ are broken. It is clear seen that their bond-lengths drastically change and become 2.859 Å. The calculated EFL shows there are no electrons to localize between B atoms. B-B bonds are thus fatal for the mechanical property of $FeB_4$, and these broken B-B bonds can result in structural collapse. The bond angles between different bonds are also shown in Fig. 2. It is noted that there are no bonds aligned in the < 111> tensile direction, and the bond angle relaxation plays an important role in the overall structural response. It is clearly seen that all of the angles vary with the strain at a lower rate with a big jump at the bond breaking point, and the variation of $\angle B_2B_1Fe_2$ is more obvious than others. Meanwhile, we calculated the strain energy associated with the <111> tensile deformation. $FeB_4$ has an energy barrier of 0.29 eV/atom.

It is well known that the ideal strength of materials is determined by their tensile and shear strength. In order to investigate ideal shear strength of $FeB_4$, various shear-sliding planes were systematically studied along different inequivalent directions under shear deformation, as shown in Fig. 3. It is clearly seen that the shear strength of $FeB_4$ is lower than its tensile strength, ranging from 24.6 to 46.3 GPa, where the weakest shear strength is along the $(111)[11\overline{2}]$ shear direction. Obviously, the shear deformation of $FeB_4$ is much weaker than that of diamond (96.3 GPa[19]) and c-BN (70.5 GPa[19]) along the weakest shear direction of $(111)[11\overline{2}]$ direction. To further analyze the bond-breaking mechanism of $FeB_4$ in its weakest pure shear deformation direction $(111)[11\overline{2}]$, we examine the bond length, bond angle, and the strain energy variations of $FeB_4$ along the easy-slip $(111)[11\overline{2}]$ direction, as shown in Fig. 4(left panel). At $\varepsilon = 0$, the bond-length of $Fe_1$-$B_1$ is 2.247 Å and $B_2$ is bonding with $B_3$, $B_4$, $B_5$ and $B_6$, respectively, where bond-lengths are as follows: $|B_2$-$B_3| = 1.705$ Å, $|B_2$-$B_4| = 1.879$ Å, $|B_2$-$B_5| = |B_2$-$B_6| = 1.839$ Å. As the shear deformation increases in the $(111)[11\overline{2}]$ direction, the distance of $Fe_1$-$B_1$ increases and reaches 3.251 Å at $\varepsilon = 0.42$, indicating collapse of this bond. Moreover, the bond length of $|B_2$-$B_3|$ decreases while that of $|B_2$-$B_5|$ increases, accompanying a charge transfer from $B_2$-$B_3$ bond to the center of $\Delta B_2B_3B_4$ and forming an interesting three-center bond in $FeB_4$. The bond-length of $B_2$-$B_5$ achieves 2.734 Å, however, the bonding between $B_2$ and $B_6$ atoms becomes stronger due to the shorter bond-length of 1.725 Å at $\varepsilon = 0.42$. Fig. 4 also shows the different angle variations in $FeB_4$ under the $(111)[11\overline{2}]$ shear. It is obvious that the variation of $\angle B_8B_3B_4$ is comparably bigger than others.

Fig. 5(a) and 5(b) show the band structure near the Fermi energy and electronic densities of $FeB_4$ at 0 GPa, respectively. The zero energy refers to the top of valence bands. As shown in Fig. 5(a), we note that there are several bands crossing the Fermi level, indicating the good electronic mobility in $FeB_4$. This special character of $FeB_4$ brings its special application on electron conductivity. Fig. 5(b) implies the good metallicity due to the large total DOS at the Fermi level. From the calculated partial DOS in Fig. 5(b), we can see that the electronic structure of $FeB_4$ is governed by the strong hybridization between the Fe-d and B-p states, while with a rather small contribution from the Fe-s, Fe-p and B-s states. This hybridization of Fe-d and B-p also indicates the strong interaction between Fe and B atoms. Furthermore, we have also investigated the change of electronic band structure under tensile and shear deformation; see Fig.5 (c) and (d). It clearly seen that there are obvious differences under deformation, indicating the critical bond breaking and collapse of this structure.

**CONCLUSIONS**

We have performed systematic first-principles simulations to examine the structural stability, electronic properties, Vickers hardness, tensile and shear ideal strength of $FeB_4$, which was measured by experiment to be a superhard material. Our current results reveal the bulk and shear modulus of $FeB_4$ is 265 and 198 GPa, respectively, which is much lower than those found in BN and diamond. The calculated weakest tensile and shear stresses are 40 GPa and 25 GPa, respectively, and its Vickers hardness is thus estimated to be ~25 GPa, which is below the superhardness threshold of 40 GPa. Moreover, we examined the atomistic bonding structural variations to elucidate the microscopic mechanism for obtained stress-strain relation of $FeB_4$, providing a detailed description of its structural property under the tensile and shear deformation. The present results suggest $FeB_4$ is unlikely to become superhard, with hardness exceeding 40 GPa. The emerging mutual disagreements between recent experiment and our theory are so encouraging that further experimental and theoretical studies on $FeB_4$ will be greatly stimulated.

**ACKNOWLEDGEMENTS**

This work is supported by the Natural Science Foundation of China under grants

11304167, 51374132, 11304141, 11304140, and 51202084. The authors acknowledge the High Performance Computing Center of Jilin University for supercomputer time.

*Author to whom correspondence should be addressed: zhangmiaolmc@126.com or hal420@mail.usask.ca

TABLE I. The calculated constants $a$, $b$, and $c$ (Å), bulk modulus ($B_0$), shear modulus ($G$), Young's modulus ($Y$), Poisson's ratio $v$, and Vickers hardness ($H_v$) with different empirical models, as well elastic constants $C_{ij}$ (GPa) for FeB$_4$ compared with previous calculations and available experimental results.

| | $a$ | $b$ | $c$ | $B_0$ | $G$ | $Y$ | $v$ | $H_v^{\text{Tian}}$ | $H_v^{\text{Chen}}$ | $H_v^{\text{Tsc}}$ |
|---|---|---|---|---|---|---|---|---|---|---|
| FeB$_4$ | 4.579 | 5.298 | 2.999 | 264.73 | 197.97 | 475.41 | 0.2 | 11.7 | 28.4 | 32.3 |
| | 4.521 | 5.284 | 3.006[a] | | | | | | | |
| Expt. | 4.5786 | 5.2981 | 2.9991 | 252(5) | | | | 62(5)[b] | | |
| Diamond | | | | 467.4 | 550.1 | 1185.3 | 0.0077 | 97.5 | 94.0 | 88.1 |
| Expt. | | | | 443[c] | 525.5[c] | | | 96±5[d] | | |
| c-BN | | | | 403.0 | 411.5 | 921.0 | 0.119 | 72.0 | 66.3 | 66.1 |
| Expt. | | | | 368[c] | 397.5[c] | | | 63±5[d] | | |
| | $C_{11}$ | $C_{22}$ | $C_{33}$ | $C_{44}$ | $C_{55}$ | $C_{66}$ | $C_{12}$ | $C_{13}$ | $C_{23}$ | |
| | 432.50 | 753.40 | 455.60 | 223.50 | 156.26 | 221.98 | 138.56 | 133.15 | 130.21 | |

[a]Reference 36.
[b]Reference 23.
[c]Reference 38.
[d]Reference 39.

**Figure captions**

**Fig. 1 (color online)** The calculated stress-strain relations of $FeB_4$ in various tension deformation directions.

**Fig. 2 (color online)** Left panel: The calculated bond lengths, the bond angles, and the corresponding strain energy of $FeB_4$ under the <111> tensile direction. Right panel: The structure of $FeB_4$ at equilibrium ($\varepsilon = 0$) with the three-dimensional ELF isosurfaces at ELF = 0.75.

**Fig. 3 (color online)** The calculated stress-strain curves on various shear-sliding planes in different directions under shear deformation for $FeB_4$.

**Fig. 4 (color online)** Left panel: The calculated bond lengths, the bond angles, and the corresponding strain energy of $FeB_4$ under the $(111)[11\overline{2}]$ shear direction. Right panel: The calculated structural snapshots of $FeB_4$ at equilibrium ($\varepsilon = 0$) (top) and in the $(111)[11\overline{2}]$ shear direction under shear $\varepsilon = 0.42$ (bottom), with the three-dimensional ELF isosurfaces at ELF = 0.75.

**Fig. 5 (color online)** (a) Calculated electronic band structure and (b) electronic densities of $FeB_4$ at 0 GPa. The zero energy refers to the top of valence bands. (c) Calculated electronic band structures in the <111> tensile direction at $\varepsilon = 0.23$ (blue line) and $\varepsilon = 0.24$ (red line). (d) Calculated electronic band structures in the $(111)[11\overline{2}]$ shear direction at $\varepsilon = 0.42$ (blue line) and $\varepsilon = 0.44$ (red line).

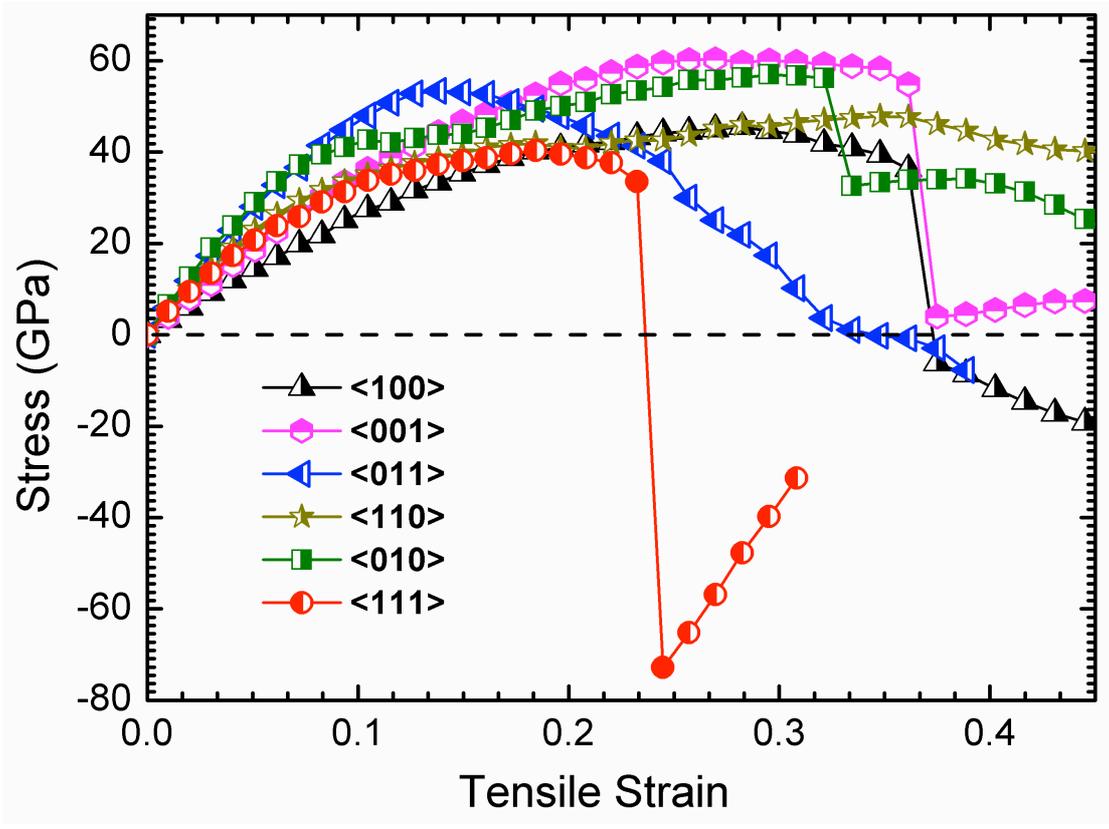

**Fig. 1**

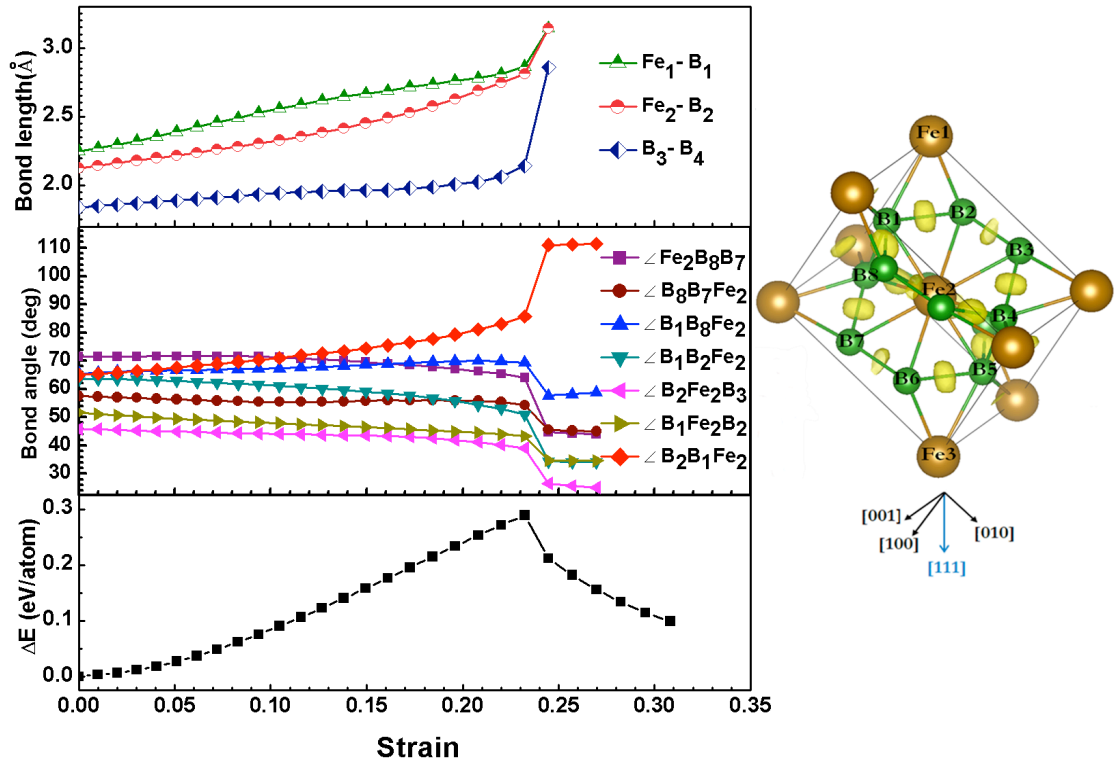

**Fig. 2**

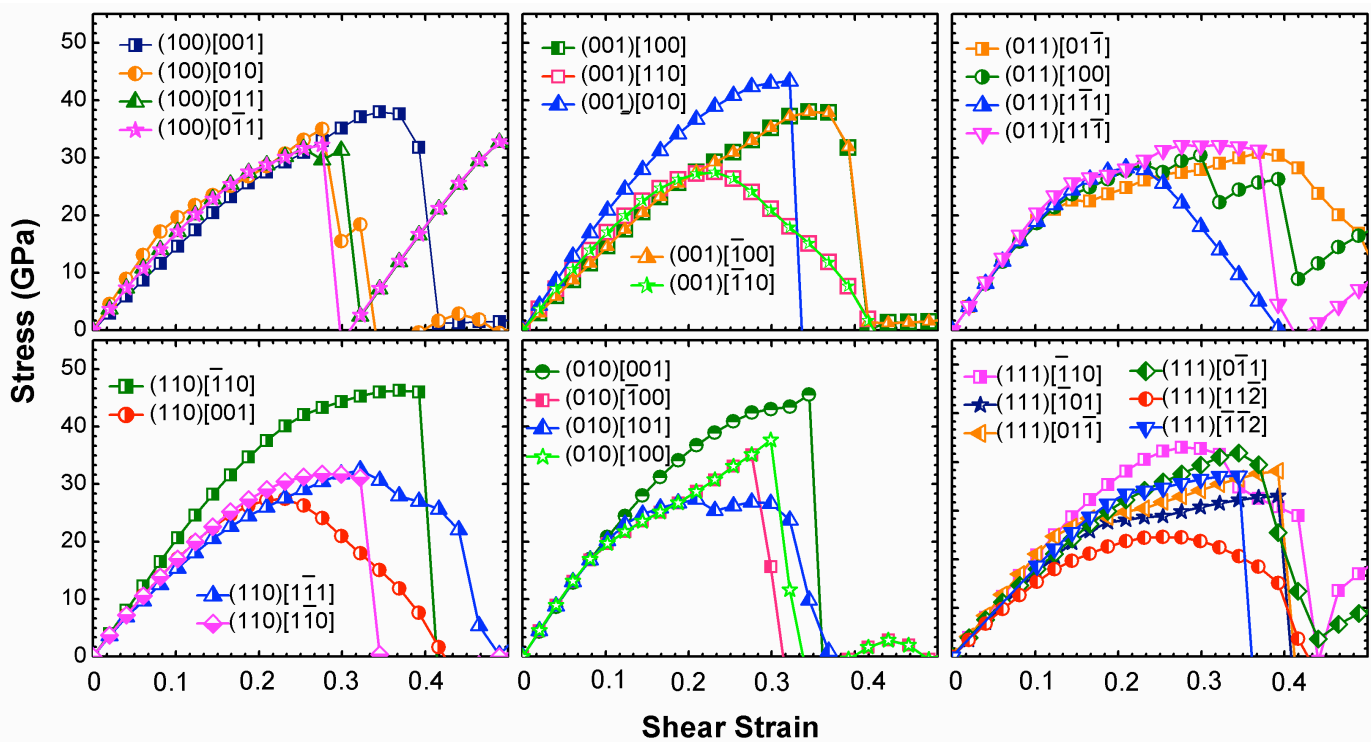

**Fig. 3**

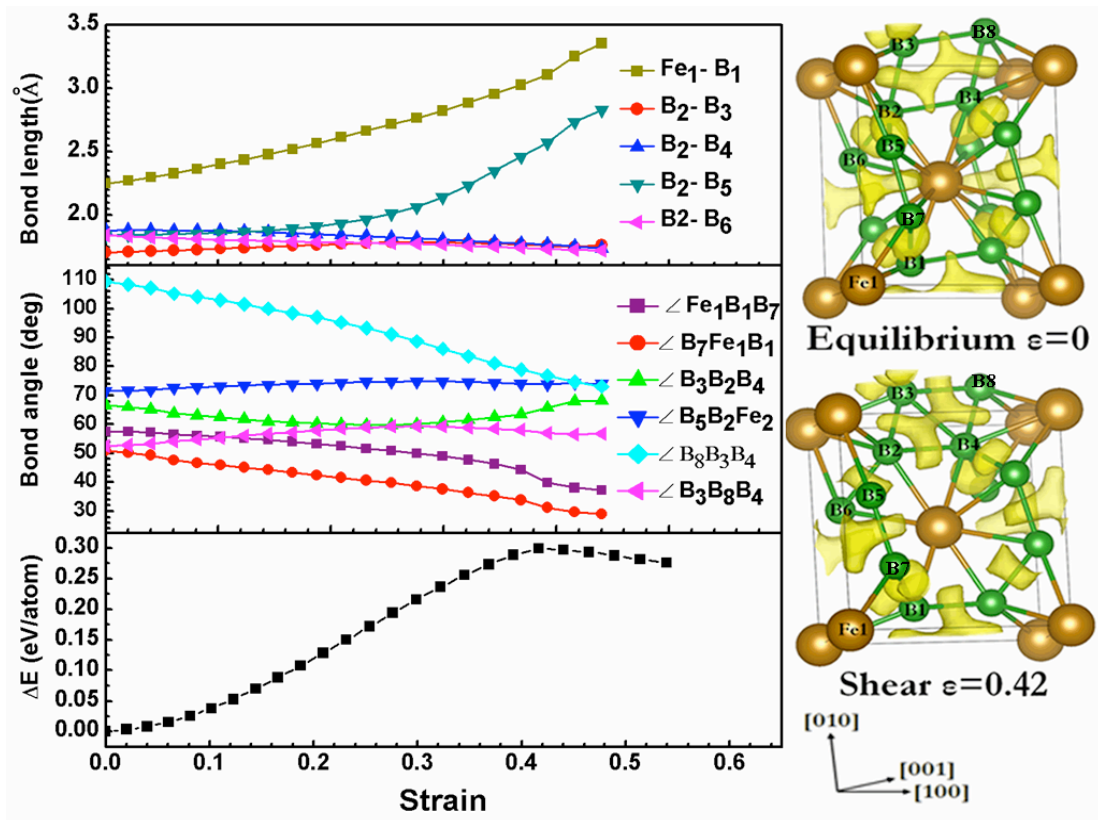

**Fig. 4**

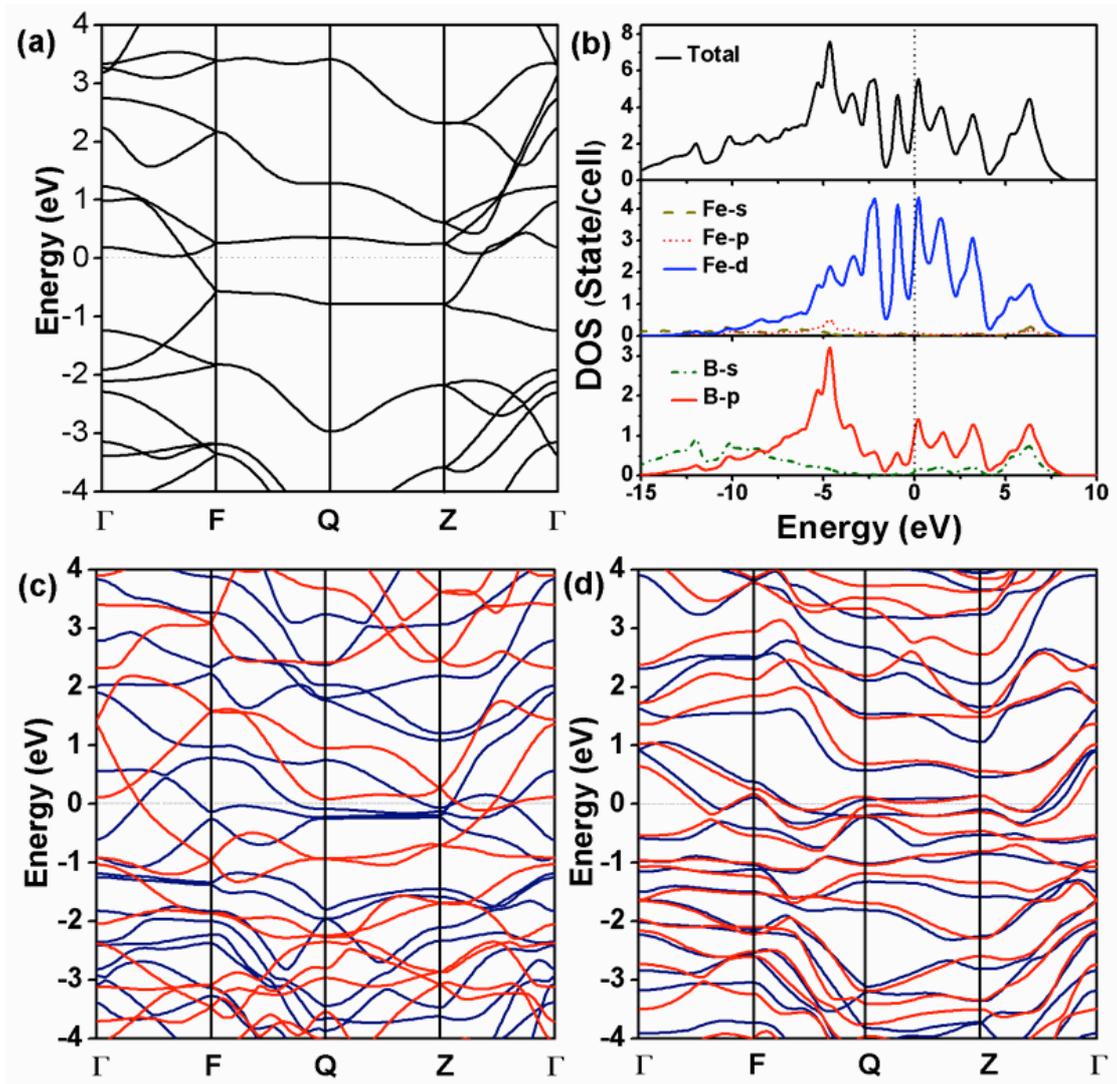

**Fig. 5**